\documentclass[acmsmall]{acmart}
\usepackage{caption,subcaption}
\usepackage{pdflscape}
\usepackage{afterpage}
\usepackage{longtable}
\usepackage{makecell}
\usepackage[normalem]{ulem}
\usepackage[utf8]{inputenc}
\usepackage{listings}
\usepackage{listings-rust}

\usepackage{multirow}
\usepackage{filecontents}
\usepackage{framed}
\usepackage{tikz} 
\usetikzlibrary{plotmarks}
\acmConference[Preprint]{This is a pre-print version}{1 Feb, 2021}

\begin{document}

\title{Memory-Safety Challenge Considered Solved? An In-Depth Study with All Rust CVEs}

\author{Hui Xu}
 \affiliation{%
   \institution{School of Computer Science, Fudan University}
   }

\author{Zhuangbin Chen}
 \affiliation{%
   \institution{Dept. of CSE, The Chinese University of Hong Kong}
 }

 \author{Mingshen Sun}
 \affiliation{%
   \institution{Baidu Security}
 }

\author{Yangfan Zhou}
 \affiliation{%
   \institution{School of Computer Science, Fudan University}
   }

\author{Michael R. Lyu}
 \affiliation{%
   \institution{Dept. of CSE, The Chinese University of Hong Kong}
   }

\begin{abstract}
Rust is an emerging programing language that aims at preventing memory-safety bugs without sacrificing much efficiency. The claimed property is very attractive to developers, and many projects start using the language. However, can Rust achieve the memory-safety promise? This paper studies the question by surveying 186 real-world bug reports collected from several origins which contain all existing Rust CVEs (common vulnerability and exposures) of memory-safety issues by 2020-12-31. We manually analyze each bug and extract their culprit patterns. Our analysis result shows that Rust can keep its promise that all memory-safety bugs require unsafe code, and many memory-safety bugs in our dataset are mild soundness issues that only leave a possibility to write memory-safety bugs without unsafe code. Furthermore, we summarize three typical categories of memory-safety bugs, including automatic memory reclaim, unsound function, and unsound generic or trait. While automatic memory claim bugs are related to the side effect of Rust newly-adopted ownership-based resource management scheme, unsound function reveals the essential challenge of Rust development for avoiding unsound code, and unsound generic or trait intensifies the risk of introducing unsoundness. Based on these findings, we propose two promising directions towards improving the security of Rust development, including several best practices of using specific APIs and methods to detect particular bugs involving unsafe code. Our work intends to raise more discussions regarding the memory-safety issues of Rust and facilitate the maturity of the language.
\end{abstract}

\maketitle

\pagenumbering{arabic}
\pagestyle{plain}

\section{Introduction}
Memory-safety bugs (\textit{e.g.,} buffer overflow) are critical software reliability issues~\cite{szekeres2013sok}. Such bugs commonly exist in software written by system programming languages like C/C++ that allow arbitrary memory access. The problem is challenging because regulating memory access while not sacrificing the usability and efficiency of the language is difficult. Rust is such a system programming language that aims at addressing the problem. Since the release of its stable version in 2015, the community of Rust grows very fast, and many popular projects are developed with Rust, such as the web browser Servo~\cite{anderson2016engineering}, and the operating system TockOS~\cite{levy2017tock}.  

To achieve memory-safety programming, Rust introduces a set of strict semantic rules for writing compiling code and therefore preventing bugs. The core of these rules is an ownership-based resource management (OBRM) model, which introduces an owner to each value. Ownership can be borrowed among variables as references or aliases in two modes: mutable or immutable. The main principle is that a value cannot have multiple mutable aliases at one program point. Once a value is no longer owned by any variable, it would be dropped immediately and automatically. In order to be more flexible and extensible, Rust also supports unsafe code which may break the principle but should be denoted with an \texttt{unsafe} marker, meaning that developers should be responsible for the risk of using such code. Existing memory-safety checking mechanisms enforced by Rust compiler are unsound for unsafe code. 

As Rust surges into popularity, a critical concern to the software community is how Rust performs in combating memory-safety bugs. Previous to our work, Evans \textit{et al.}~\cite{Evans2020is} have performed a large-scale study showing that unsafe code is widely used in Rust crates (projects). However, it remains unclear whether and how unsafe code could undermine the memory safety of real-world Rust programs. Since existing materials lack an in-depth understanding, we attempt to address this question by empirically investigating a set of critical bugs reported in real-world Rust projects. When we are preparing this work, we notice another independent work~\cite{qin2020understanding} also studies the same problem. While their work has developed some understandings similar to our paper, our analysis involves more memory-safety bugs and provides several novel findings different from them, such as the culprit patterns of using generic or trait. We will clarify the differences in Section~\ref{sec:diff}.

To elaborate, our work collects a dataset of memory-safety bugs from several origins, including Advisory-DB, Trophy Case, and several open-source projects on GitHub. The dataset contains 186 memory-safety bugs in total, and it is a superset of all Rust CVEs with memory-safety issues by 2020-12-31. For each bug, we manually analyze the consequence and culprit and then classify them into different groups. Since the consequences are nothing special in comparison with those of traditional C/C++~\cite{szekeres2013sok}, we adopt a top-down approach and directly employ the labels like buffer overflow/over-read, use-after-free, and double free. However, we have no prior knowledge regarding the patterns of culprits in Rust, so we adopt a bottom-up approach and group them if two culprits demonstrate similar patterns.

Based on these bugs, this paper studies three major questions. The first question is how effective is Rust in preventing memory-safety bugs. Our inspection result shows that all bugs in our dataset require unsafe code except one compiler bug. The main magic lies in that Rust does not tolerate unsound APIs provided by either its standard library (std-lib) or third-party libraries. As a result, many memory-safety bugs are mild issues merely introducing unsound APIs, and they cannot accomplish memory-safety crimes by themselves. Therefore, we can conjecture that Rust has kept its promise that developers are not able to write memory-safety bugs without using unsafe code.

Our second question concerns the typical characteristics of memory-safety bugs in Rust. To this end, we have extracted three typical categories of memory-safety bugs, which are automatic memory reclaim, unsound function, and unsound generic or trait. Auto memory reclaim bugs are related to the side effects of Rust OBRM. Specifically, Rust compiler enforces automatic destruction of unused values, \textit{i.e.,} by calling \texttt{drop(v)} where \texttt{v} outlives its usefulness. The design is supposed to mitigate the memory leakage issue because Rust has no runtime garbage collection, but it leads to many use-after-free, double free, and freeing invalid pointer (due to uninitialized memory access) issues in our dataset. The default stack unwinding strategy of Rust further intensifies such side effects. The category of unsound function manifests the essential challenge for Rust developers to avoid unsoundness issues in their code. The category of unsound generic or bound reveals another essential challenge faced by Rust developers for delivering sound APIs when employing sophisticated language features like polymorphism and inheritance. Besides, there are also some bugs that do not belong to the previous categories, and they are mainly caused by common mistakes, such as arithmetic overflow or boundary checking issues, which are not our interest.

Our third question studies the lessons towards making Rust programs more secure. On one hand, we find several common patterns of bugs and patches that can be employed as the best practices for code suggestion, such as bounding the generic-typed parameters with \texttt{Send} or \texttt{Sync} trait when implementing \texttt{Send} or \texttt{Sync} trait, or disabling automatic drop (with \texttt{ManuallyDrop<T>}) as earlier as possible. On the other hand, we may extend the current static analysis scheme of Rust compiler to support unsafe code so as to detect some specific types of bugs, especially those related to automatic memory reclaim. 

We highlight the contribution of our paper as follows. 

\begin{itemize}

\item This paper serves as a pilot empirical study of Rust in preventing memory-safety bugs. It performs an in-depth analysis regarding the culprits of real-world memory-safety bugs and generates several major categories, including automatic memory reclaim, unsound function, unsound generic or trait, and other errors.

\item We highlight that a unique challenge of Rust development is to deliver sound APIs, and using advanced features of Rust (\textit{e.g.,} generic or trait) intensifies the risks of introducing unsoundness issues. Furthermore, we underline that automatic memory reclaim bugs are related to the side effect of Rust newly-adopted OBRM, and the default stack unwinding strategy of Rust further exacerbates such problems.

\item Our work proposes two promising directions towards improving the security of Rust development. To developers, we provide several best practices for helping them writing sound code. To compiler/tool developers, our suggestion is to consider extending the current safety checking to the realm of unsafe code, especially in detecting the bugs related to automatic memory reclaim. 

\end{itemize}

The rest of the paper is organized as follows. Section~\ref{sec:preliminary} reviews the memory-safety war of Rust; Section~\ref{sec:approach} introduces our study approach; Section~\ref{sec:bugoverview} overviews the characteristics of our collected bugs; Section~\ref{sec:patterns} demonstrates the detailed categories and patterns of these bugs. Section~\ref{sec:lesson} discusses the lessons learned; Section~\ref{sec:threat} clarifies the threats of validity; Section~\ref{sec:related} presents related work; finally, Section~\ref{sec:conclusion} concludes the paper.

\section{Preliminary}\label{sec:preliminary}
This section reviews the preliminary of memory-safety bugs and discusses the mechanisms of Rust in preventing them.

\subsection{Memory-Safety Bugs}\label{sec:pre-memsafebugs}
Memory-safety bugs are serious issues for software systems. According to the statistical report of MITRE\footnote{https://cwe.mitre.org/top25}, memory-safety bugs are enumerated among the top dangerous software vulnerabilities. Next, we review the concept of memory-safety bugs and discuss why preventing them is hard.

\subsubsection{Concept of Memory-Safety Bugs}
In general, memory-safety bugs are caused by arbitrary memory access. Below, we discuss several typical types of memory-safety issues.

\begin{itemize}
\item \textit{Use-after-free}: After freeing a pointer, the buffer pointed by the pointer would be deallocated or recycled. However, the pointer still points to the deallocated memory address, known as a dangling pointer. Dereferencing a dangling pointer causes \textit{use-after-free} issues. Such bugs are dangerous because the freed memory may have already been reallocated for other purposes, or it may allow users to control the linked list of free memory blocks maintained by the system (\textit{e,g.,} with ptmalloc or tcmalloc~\textit{ferreira2011experimental}) and achieve arbitrary write~\cite{wu2018fuze}.

\item \textit{Double free}: Freeing a pointer twice causes \textit{double free} issues. Similar to use-after-free on heap, double free leaves room for attackers to manipulate the linked list of free memory blocks. Consequently, attackers may overwrite any memory slot (\textit{e.g.,} global offset table) of the process leveraging memory allocation system calls~\cite{eckert2018heaphopper}. 

\item \textit{Buffer overflow/over-read}: Accessing a memory slot beyond the allocated boundary is undefined. It could happen in two situations: an out-of-bound offset (\textit{e.g.,} due to boundary check error) or an in-bound offset with an invalid unit size (\textit{e.g.,} due to type conversion or memory alignment issue).

\item \textit{Uninitialized memory access}: It means the memory is used without proper initialization. Intuitively, it may leak the old content. If the memory contains values of pointer types, falsely accessing or free the pointer would cause \textit{invalid memory access}.
\end{itemize}

Note that some \textit{concurrency bugs} may also incur memory-safety issues due to data race or race conditions. The consequences could be use-after-free, buffer overflow, or \textit{etc}. However, not all concurrency bugs involve memory-safety issues, such as deadlock\footnote{https://github.com/diem/diem/pull/4479}, or logical errors that simply crash the program without segmentation faults\footnote{https://github.com/diem/diem/pull/5190}. Therefore, this work does not treat concurrency issues as a sub-type of memory-safety bugs but another different bug type orthogonal to memory-safety issues.

\subsubsection{Challenges in Preventing Memory-Safety Bugs}
As long as developers use pointers, eradicating memory-safety bugs is very difficult if not impossible. Classical strategies to combat these bugs are mainly two-fold. One is to design memory-safety protection mechanisms, such as stack canary and shadow stacks \cite{dang2015performance} for protecting stack integrity, ASLR~\cite{evtyushkin2016jump} against code reuse, glibc fasttop for detecting double free, \textit{etc}. Although such mechanisms are effective, they only raise the bar of attacks but cannot prevent them~\cite{gras2017aslr}. The other strategy is to prevent introducing memory-safety bugs from the beginning, such as using type-safe languages~\cite{ryder2005impact} and banning raw pointers, like Java. However, the enforcement of the memory-safety feature may also pose limitations to the language, making it inefficient for system-level software development with rigorous performance requirements.

\subsection{Memory-Safety of Rust}

\begin{figure}[t!b]
\includegraphics[width=0.9\textwidth]{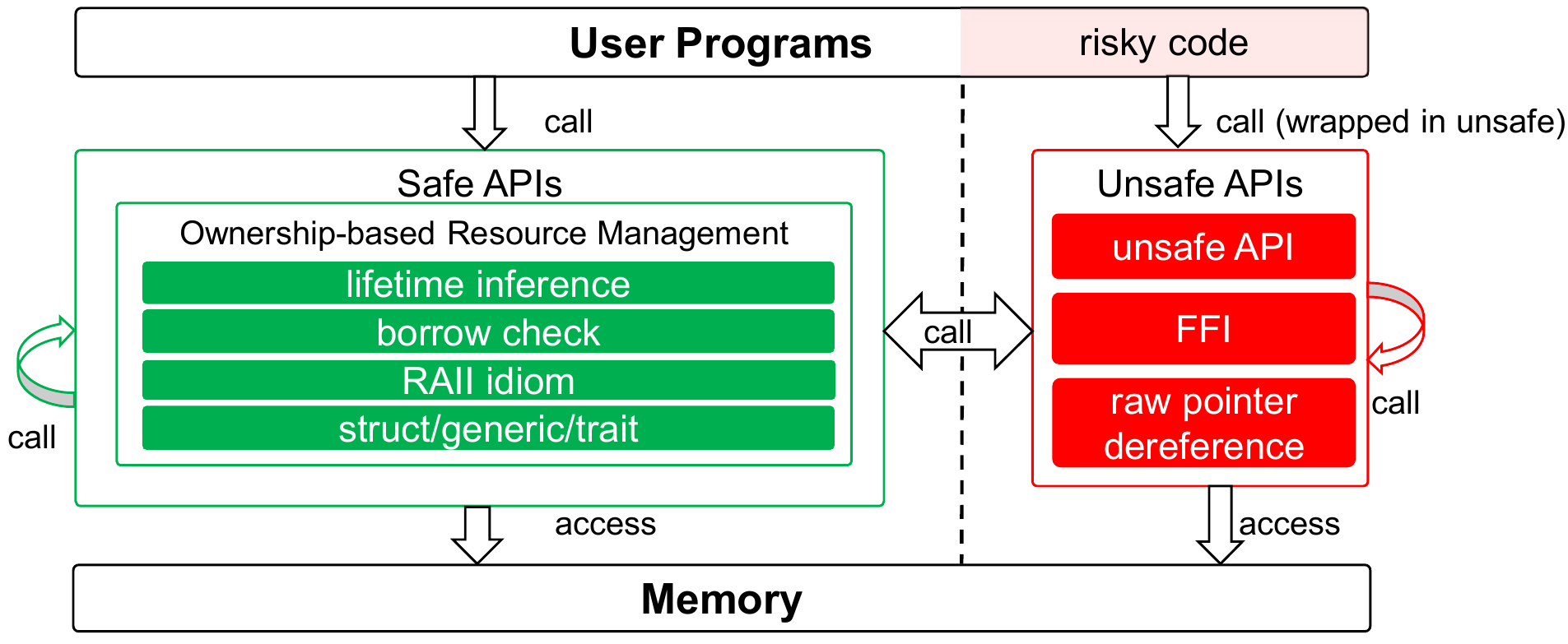}
\caption{Idea of Rust for preventing memory-safety bugs.}
\label{fig:unsafe}
\end{figure}

Rust is a system programming language that aims at preventing memory-safety bugs while not sacrificing performance. It approaches the goal by introducing a set of strict semantic rules at the compiler level. In this way, Rust can be more efficient than other programming languages (\textit{e.g.,} Go) that rely much on runtime memory checking and garbage collection~\cite{cutler2018benefits}. 

\subsubsection{Basic Design} Figure~\ref{fig:unsafe} overviews the idea of Rust. Rust is in nature a hybrid programming language, including a safe part and an unsafe part. The safe part guarantees that the behaviors of all the code and APIs are well-defined, and programs using safe APIs only should have no risk of memory-safety issues. The unsafe part has no such guarantee and may lead to undefined behaviors, but it is necessary to meet some specific needs, \textit{e.g.,} efficient memory access for software development with rigorous performance requirements. Any code that may lead to undefined behaviors should be declared as unsafe, such as dereferencing raw pointers, calling FFIs (foreign function interfaces), and unsafe APIs. Actually, many safe APIs also employ unsafe APIs internally, and these APIs can be safe because they have got rid of all memory-safety risks, \textit{e.g.,} via conditional code. A function calling unsafe APIs can be declared as either safe or unsafe, which mainly depends on the developer's decision. Rust cannot check whether the declaration is correct or not. Therefore, falsely declaring an API as safe is dangerous and could undermine the soundness of safe Rust. Besides, Rust is not interested in memory leakage issues. Any code that may cause memory leakage is safe in Rust. Therefore, our following discussion will not include memory leakage issues.

The core of safe Rust is a novel ownership-based resource management model~\cite{jung2017rustbelt}. Ownership means a value should have one variable or identifier as its owner, and ownership is exclusive that a value can have only one owner. However, ownership can be borrowed among variables as references or aliases in two modes: immutable (default) or mutable (with an extra \texttt{mut} marker). The model assumes that only one variable can have mutable access to a value at any program point while other aliases have neither mutable nor immutable access at that point, or only immutable access can be shared among multiple aliases. Rust implements a \textit{borrow checker} in its compiler to achieve the goal. The borrowed ownership expires automatically when the program execution exits the scope. If a value is no longer owned by variables, it would be dropped immediately to reclaim the buffer no longer used. Associated with the model, Rust introduces a \textit{lifetime inference} mechanism (similar as type inference~\cite{odersky1999type}) which assures that the lifetime of a borrowed ownership would last long enough for use. Together, they form a basis for Rust in preventing memory-safety bugs. The realization of ownership-based resource management is based on the RAII (resource acquisition is initialization) idiom. RAII emphasizes resource allocation during object creation with the constructor and resource deallocation during object destruction with the destructor~\cite{stroustrup2001exception,ramananandro2012mechanized}. Therefore, the borrow-check and lifetime-inference algorithms only need to consider well-initialized objects and their references. In other words, OBRM saves Rust compiler from solving complex pointer analysis problems and thus largely simplifies the game played by the compiler. 

The OBRM model applies to all types supported by Rust, including self-defined structs, generics, and traits. Generic relates to the polymorphism feature of Rust which means an unspecified type (often denoted with \texttt{T}) that will be determined during compilation (similar to C++ template). Trait relates to the inheritance feature of Rust which means a trait type can be inherited and its member functions can be reimplemented by the derived types, such as the \texttt{Clone} trait and \texttt{Drop} trait. When implementing Rust code, a generic can be bounded with some traits to be sound. 

\subsubsection{Principle for Preventing Memory-Safety Bugs}
OBRM lays the foundation of Rust in preventing memory-safety bugs. Via OBRM and other related design, Rust compiler assures Rust developers that they will not suffer memory-safety issues if not using unsafe code. Below, we justify the effectiveness of OBRM in preventing different memory-safety issues.

\begin{itemize}
\item \textit{Preventing dangling pointer}: According to the OBRM, variables or pointers should be initiated with values when defining them. This enables the compiler to trace the abstract states (ownership and lifetime) of the value and perform sound program analysis. Such analysis guarantees that safe Rust can prohibit shared mutable aliases, and therefore gets rid of the risks of dereferencing or freeing pointers whose values have been freed with another alias. Note that while defining raw pointers is valid in safe Rust, dereferencing them is only allowed as unsafe. However, the soundness of Rust compiler would be invalidated when using unsafe code. For example, unsafe code could introduce shared mutable aliases and therefore makes the program vulnerable to double free due to the automatic memory reclaim scheme. We will elaborate more on this problem in Section~\ref{sec:autoreclaim}.

\item \textit{Preventing buffer overflow/over-read}: OBRM also benefits in-range buffer access because each variable or pointer points to a value of a particular type, such as \texttt{i32}. It guarantees that memory data are properly aligned. For advanced data containers of Rust std-lib, such as \texttt{Vec<T>}, Rust generally maintains a length field for the object and performs boundary checks automatically during runtime. Again, these mechanisms are only sound for safe Rust, unsafe code (such as unsafe type conversion or dereferencing raw pointers) may lead to out-of-range access.

\item \textit{Preventing uninitialized memory access}: Safe Rust does not allow uninitialized memories by default. For example, creating a buffer with \texttt{uninitialized()} or \texttt{alloc()} is unsafe; reserving the capacity for a vector is safe, but directly increasing its length with \texttt{set\_len()} is unsafe.
\end{itemize}

OBRM also applies to concurrent programming and therefore can mitigate memory-safety risks incurred by data race. For example, when accessing a variable defined in the main thread from a child thread, a marker \texttt{move} is needed to transfer the ownership of the variable. To access variables concurrently in multiple processes, Rust provides \texttt{Mutex<T>} and \texttt{Arc<T>} that achieve mutual exclusive lock and atomic access respectively, and the type system of Rust ensures that ownerships cannot be shared among threads unless they implementing \texttt{Send} or \texttt{Sync} traits, \textit{e.g.,} based on \texttt{Mutex<T>} or \texttt{Arc<T>}. These features can help developers avoid introducing data race or race conditions in their code.

Another essential trick of Rust is that safe Rust does not tolerate unsoundness, \textit{i.e.,} if a code snippet provides some safe APIs but which can be employed by developers to accomplish undefined behaviors without using unsafe code, the code snippet contains unsoundness. Rust requires that all safe APIs should not be able to incur undefined behaviors, otherwise it should be unsafe. This requirement applies to all Rust projects and developers, and it plays an important role as the Rust community gradually grows and many third-party libraries become available. Only when all these libraries are sound, the soundness of Rust compiler can be guaranteed.

\section{Our Approach}\label{sec:approach}
\subsection{Objective}
This work aims at reviewing the memory-safety game played by Rust. In particular, we are interested in three major research questions.

\uline{\textit{RQ-1. How effective is Rust in preventing memory-safety bugs?}} Rust promises developers that writing programs without using unsafe code should have no memory-safety risks. Therefore, we will study whether unsafe code is necessary for accomplishing real-world memory-safety bugs and how bad are these residual bugs. Answering these questions provides an overall assessment of Rust in combating memory-safety issues, which is especially useful to its potential users.

\uline{\textit{RQ-2. What are the characteristics of memory-safety bugs in Rust?}} For those residual memory-safety bugs in real-world Rust programs, what are their causal culprits? Are there any unique patterns in comparison with C/C++ bugs?  Interpreting such issues can provide more insights towards understanding the problem better or mitigating them in the future.

\uline{\textit{RQ-3. What lessons can we learn in order to make Rust more secure?}} Based on the findings of RQ1 and RQ2, we shall be able to derive some useful knowledge, helping Rust developers further mitigate residual memory-safety issues. For example, we can extract some common code patterns that are risky to memory-safety issues and warn developers to use them carefully. Likewise, the characteristics of bugs may provide essential information for compiler developers to consider improving the language towards more resilience. 

\subsection{Data Collection}
The first step of our study is to collect real-world memory-safety bugs. Thanks to GitHub, many influential Rust projects are maintained online and allow us to track the bugs and patches for particular projects. However, since GitHub is free-organized, the formats and qualities of bug reports vary a lot. Such limitations pose difficulties for automatically extracting and analyzing bugs in-depth. As our research does not focus on proposing automatic analysis approaches, we prefer manual inspection which can be performed in an in-depth and precise manner. While manual inspection can be in-depth, it sacrifices scalability. Therefore, we try to find the most representative cases for manual inspection. 

We collect bugs from four origins, Advisory-DB, Trophy Cases, official Rust compiler project, and several other Rust projects on GitHub. 

\begin{itemize}
\item \textit{Advisory-DB}\footnote{https://github.com/RustSec/advisory-db} is a repository that collects security vulnerabilities found in Rust software, and it is a superset of existing CVEs\footnote{https://cve.mitre.org/cgi-bin/cvekey.cgi?keyword=rust} related to Rust programs. The dataset contains over 200 entries from 100+ contributors by 2020-12-31. However, not all entries are memory-safety bugs, such as those labeled with unmaintained, crypto-failure, or denial-of-service. We finally obtained 137 memory-safety bugs (including 122 CVEs) for further study. Note that the bug selection process is via manual inspection because the original entries of Advisory-DB are provided by different developers, and they may employ different labels.

\item \textit{Trophy Case}\footnote{https://github.com/rust-fuzz/trophy-case} collects a list of software bugs found by fuzzing tools. The bug reports originate from the users of fuzzing tools for Rust programs, such as cargo-fuzz\footnote{https://github.com/rust-fuzz/cargo-fuzz}. The repository contains over 200 bug entries from 30+ contributors by 2020-12-31. Unfortunately, most of these bugs do not involve memory-safety issues, and only six are labeled with a security tag, including two duplicated ones with CVE-IDs. 

\item \textit{Rust Compiler Project} maintained on GitHub also contains some bugs that may lead to undefined behaviors. In particular, we are interested in the unsoundness bugs introduced in Rust standard libraries. To this end, we search GitHub issues with the label ``I-unsound'' and ``T-libs'' or ``impl-libs'' and inspect each of them. Note that we mainly keep closed issues for further study because they can provide us thorough information about the bug and its fix. We finally obtained 33 memory-safety bugs in Rust std-lib, including five duplicated ones with Advisory-DB (three have CVE-IDs).

\item \textit{Other Projects} To further extend our dataset, we choose five highly rated (over 500 stars) Rust projects on GitHub. We search the keywords related to memory-safety (\textit{e.g.,} ``core dumped'', ``segmentation fault'', ``use-after-free'') for finding the historical bugs occurred in these repositories. We finally obtained 18 bugs in these projects, including one duplicated CVE. Note that since the bugs of previous origins are mainly library bugs, we especially choose three projects of executables.

\end{itemize}

\begin{figure}[t!b]
\includegraphics[width=0.9\textwidth]{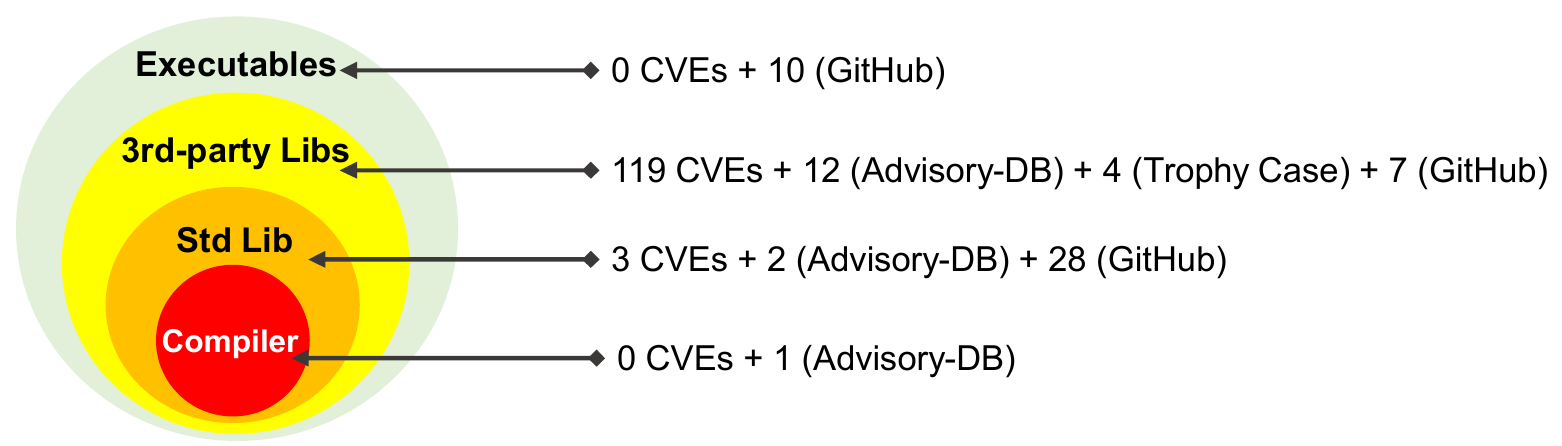}
\caption{Dataset of memory-safety bugs in our study.}
\label{fig:dataset}
\end{figure}

We finally get 186 memory-safety bugs for further study. Figure~\ref{fig:dataset} provides an overview of these bugs according to the types of their host programs, one bug in Rust compiler, 33 in Rust standard library, 142 in third-party libraries, and 10 in executable programs. A full list of these bugs and their detailed information are available online\footnote{https://github.com/Artisan-Lab/Rust-memory-safety-bugs}. 

Since Rust does not consider memory leakage as memory-safety bugs, we exclude these bugs from our dataset. In particular, many CVEs with memory leakage issues lead to DoS but not memory-safety problems. For example, a malformed input or unexpected parameter value could cause the system to preallocate a huge memory space, such as CVE-2018-20993 and CVE-2019-15544. Besides, employing recursive function calls may overflow the stack (\textit{e.g.,} 8MB on Linux X86\_64 by default) if the recursion is too deep, such as CVE-2019-2500. Such bugs mainly lead to DoS because the operating system can detect such issues and kill the process. Besides, some other memory bugs might also be confused with memory-safety ones, but they do not incur security issues. Typically, a bug with ``index out-of-bound'' error indicates the violation of index validity check performed by the container (\textit{e.g.,} a vector) itself during runtime. There are many such cases founded by fuzzing tools, like those reported by Trophy Cases\footnote{https://github.com/rust-fuzz/trophy-case}. Similar to memory leakage, these bugs can affect the usability of the program but have no security risks. We will not study these bugs in our paper.

\subsection{Bug Analysis Method}\label{sec:analysis}
We are interested in the consequences and culprits of each memory-safety bug. For the consequences, we adopt a \textit{top-down approach}. Because the taxonomy of memory-safety bugs has been well studied in the literature (\textit{e.g.,}~\cite{szekeres2013sok}), we simply adopt the existing categorization method, such as buffer overflow/over-read, use-after-free, double free, and uninitialized memory access. For the culprits, we employ a \textit{bottom-up approach}, \textit{i.e.,} we cluster similar bugs while do not assume any prior categories. While the bug description can provide us some hints, we also manually check the commit of each bug fix on GitHub and analyze their culprits. The analysis is mainly done by the first author and checked by the second and third authors. By comparing the buggy code and bug fixes, we can locate the root cause of each bug and confirm the culprits. In particular, many bugs, especially most CVEs employ similar descriptions, which provide a basis to form our culprit categories. For example, the description of CVE-2020-36206 ``\textit{lack of Send and Sync bounds}'' is similar to CVE-2020-36220 ``\textit{Demuxer<T> omits a required T:Send bound}'', and we cluster them into one group with label ``\textit{insufficient bound to generic}''; the description of CVE-2020-36210 ``\textit{uninitialized memory can be dropped when a panic occurs}'' is similar to CVE-2020-25795 ``insert\_from can have a memory-safety issue upon a panic'' and we cluster them into another group with label ``bad drop at cleanup block''. For bugs without clear descriptions, we will first try to categorize them into existing groups or employ a new group if their patterns are different. In this way, we finally obtain four major categories and 13 sub-categories. The details will be discussed in the next section (Section~\ref{sec:distribution}).

\section{Overview of Memory-Safety Bugs}\label{sec:bugoverview}
This section provides an overview of our bug analysis results, and then addresses RQ-1.

\subsection{Distribution of Bugs}\label{sec:distribution}

\begin{table}[]
\caption{Distribution of memory-safety bugs in Rust std-lib + CVEs + others. For simplicity, we count CVEs of Rust std-lib into Rust std-lib.}
\label{tab:bugs}
\scriptsize
\begin{tabular}{|c|c|c|c|c|c|c|c|}
\hline
 \multicolumn{2}{|c|}{} & \multicolumn{5}{c|}{\textbf{Consequence}} & \multirow{2}{*}{Total} \\ \cline{1-7} 
 \multicolumn{2}{|c|}{\textbf{Culprit}} & Buf. Over-R/W & Use-After-Free & Double Free & Uninit Mem & Other UB & \\ \hline
\multirow{2}{*}{\begin{tabular}[x]{@{}c@{}}Auto Memory\\Reclaim\end{tabular}}
	& Bad Drop at Normal Block 	  	& 0 + 0 + 0 & 1 + 9 + 6 & 0 + 2 + 1 & 0 + 2 + 0 & 0 + 1 + 0 & 22 \\ \cline{2-8} 
	& Bad Drop at Cleanup Block		& 0 + 0 + 0 & 0 + 0 + 0 & 1 + 7 + 0 & 0 + 5 + 0 & 0 + 0 + 0 & 13 \\ \hline 
\multirow{2}{*}{\begin{tabular}[x]{@{}c@{}}Unsound\\Function\end{tabular}}
	& Bad Func. Signature 			& 0 + 2 + 0 & 1 + 5 + 2 & 0 + 0 + 0 & 0 + 0 + 0 & 1 + 2 + 4 & 17  \\ \cline{2-8}
	& Unsoundness by FFI  			& 0 + 2 + 0 & 5 + 1 + 0 & 0 + 0 + 0 & 0 + 0 + 0 & 1 + 2 + 1 & 12 \\ \hline	
\multirow{3}{*}{\begin{tabular}[x]{@{}c@{}}Unsound\\Generic\\or Trait\end{tabular}}		
	& Insuff. Bound of Generic  	& 0 + 0 + 1 & 0 + 33 + 2 & 0 + 0 + 0 & 0 + 0 + 0 & 0 + 0 + 0 & 36 \\ \cline{2-8}
	& Generic Vul. to Spec. Type			& 3 + 0 + 1 & 1 + 0 + 0 & 0 + 0 + 0 & 1 + 0 + 1 & 1 + 2 + 0 & 10 \\ \cline{2-8}
	& Unsound Trait 	 			& 1 + 2 + 1 & 0 + 0 + 0 & 0 + 0 + 0 & 0 + 0 + 0 & 0 + 2 + 0 & 6 \\ \hline 
\multirow{5}{*}{\begin{tabular}[x]{@{}c@{}}Other\\Errors\end{tabular}}
	& Arithmetic Overflow 			& 3 + 1 + 0 & 1 + 0 + 0 & 0 + 0 + 0 & 0 + 0 + 0 & 0 + 0 + 0 & 5 \\ \cline{2-8}
	& Boundary Check 				& 1 + 9 + 0 & 1 + 0 + 0 & 0 + 0 + 0 & 0 + 0 + 0 & 1 + 0 + 0 & 12 \\ \cline{2-8}
	& No Spec. Case Handling 		& 2 + 2 + 1 & 0 + 0 + 0 & 0 + 0 + 0 & 0 + 0 + 0 & 2 + 1 + 1 & 9 \\ \cline{2-8}
	& Exception Handling Issue  	& 0 + 0 + 0 & 0 + 0 + 0 & 0 + 0 + 0 & 0 + 0 + 0 & 1 + 2 + 1 & 4 \\ \cline{2-8} 
	& Wrong API/Args Usage 			& 0 + 3 + 0 & 1 + 4 + 0 & 0 + 0 + 0 & 0 + 1 + 1 & 0 + 5 + 2 & 17 \\ \cline{2-8} 
	& Other Logical Errors  	    & 0 + 4 + 1 & 2 + 3 + 4 & 0 + 0 + 1 & 0 + 1 + 0 & 1 + 4 + 1 & 22 \\ \hline
\multicolumn{2}{|c|}{Total} 		& 40	& 82	& 12		& 12		& 39		& 185	\\ \hline
\end{tabular}
\end{table}

We categorize memory-safety bugs based on their consequences and culprits. Table~\ref{tab:bugs} overviews the statistics of our analysis result. The table header lists five major consequences we adopted in this paper, buffer overflow/over-read, use-after-free, double free, uninitialized memory access, and other undefined behaviors (\textit{i.e.,} memory-safety bugs that do not directly lead to the previous four consequences). We cluster the culprits of bugs into four major categories and 13 sub-categories. Each cell of the table presents the number of corresponding bugs with a particular culprit and consequence. If a bug has several different culprits or can lead to different consequences based on their usage, we only count the major culprit and consequence. For example, many use-after-free bugs may also lead to double free, but we mainly category them into use-after-free. To facilitate our following discussion, we also present the fine-grained distribution of bugs based on their origins (\textit{i.e.,} std-lib, CVEs, or others sources), separated with ``+''. 

Next, we present the high-level concepts of the four major culprit categories. While the first three ones are more or less typical for Rust, other errors are mainly common mistakes and not our interest.
\begin{itemize}
\item \textit{Automatic Memory Reclaim}: Bugs of this group imply memory-safety issues related to the automatic memory reclaim mechanism of OBRM. According to the place where the culprit drop statement happens, we divide this group into two sub-groups, \textit{bad drop at normal block} and \textit{bad drop at cleanup block}.
\item \textit{Unsound Function}: This group includes two sub-groups, \textit{bad function signature} which means developers publish an API with an incorrect function signature making the API unsound, and \textit{unsoundness by FFI} which means developers call an FFI but does not handle their behaviors properly. 
\item \textit{Unsound Generic or Trait}: Generic and trait are advanced features of Rust type system to support polymorphism and inheritance. Bugs of this group relate to the soundness issues of generics and traits, including \textit{insufficient bound of generic}, \textit{generic vulnerable to specific type}, and \textit{unsound trait}.  
\item \textit{Other Errors}: This group contains bugs that cannot be grouped into the previous three groups. These bugs are mainly due to various common mistakes, such as arithmetic overflow, boundary check, no special case handling, exception handling issue, wrong API usage, and other logical errors. When combining with unsafe code, these mistakes can also lead to memory-safety issues.
\end{itemize}

From Table~\ref{tab:bugs}, we can observe that use-after-free (82 bugs) is the most popular issue, and their major consequences include insufficient bound of generic (35 bugs), automatic memory reclaim (16 bugs), bad function signature (8 bugs), and \textit{etc}. Besides, there are 40 buffer overflow/over-read, 12 double free bugs, 12 uninitialized memory access, and 39 bugs with other undefined behaviors. Due to the automatic memory reclaim scheme, most use-after-free bugs in Rust may also lead to double free issues because the dangling pointer is likely to be freed automatically after use. We will discuss the issue later in Section~\ref{sec:patterns}.

Among all the 185 bugs (excluding one compiler bug), 35 bugs have culprits directly related to the automatic memory reclaim issue, 52 related to unsound generic or trait, 29 related to unsound functions, and 69 are due to other errors. While bugs of the first three categories are our interest because they are special in Rust, we are not interested in other errors because their patterns commonly exist in other languages. For example, arithmetic overflow and incorrect boundary check are common mistakes, and the only difference for Rust is that these bugs involve unsafe read/write to trigger memory-safety problems. 

\subsection{Effectiveness of Rust}
Based on the dataset, now we answer RQ-1. To this end, we first present our main result and then discuss the detailed analysis process by answering three sub-questions.

\uline{\textbf{Result Highlight for RQ-1}: All memory-safety bugs of Rust except compiler bugs require unsafe code. The main magic lies in that Rust enforces the soundness requirement on APIs provided by both Rust standard library or third-party libraries. As a result, many bugs in our dataset are mild soundness issues and these bugs have not incurred severe memory-safety problems. Besides, the trend of compiler bugs is much stable. Therefore, we can conclude the effectiveness of Rust in preventing memory-safety bugs.}

\uline{\textit{RQ-1.1 Do all memory-safety bugs require unsafe code?}}

According to the memory-safety promise of Rust, developers should not be able to write Rust code that contains memory-safety bugs if not using unsafe Rust. Therefore, we first examine if all these bugs require unsafe APIs. Our result confirms that all memory safety-bugs in our dataset except the compiler bug require unsafe code. According to Figure~\ref{fig:dataset}, a bug may propagate from an inner circle to outer circles, \textit{i.e.,} compiler -> std-lib -> third-party lib -> executable. For example, an unsoundness issue of the Rust compiler generally implies accepting ill-formed source code or generating incorrect compiled code, and it may affect any programs being compiled, including std-lib. Our dataset contains only one compiler bug that accepts source code violating the lifetime rule of Rust. Consequently, developers may safely extend a bounded-lifetime value to static, leading to use-after-free bugs in the compiled code. This bug is the only compiler bug collected by Advisory-DB but not the sole one ever introduced to Rust compiler. For other memory-safety issues, they could be traced back to either its own unsafe code or other unsafe code of its required libraries. For example, a memory-safety issue of an executable program that forbids unsafe code could be caused by the unsoundness of std-lib. 

As there are still many memory-safety bugs in Rust projects, we are interested in how severe are these bugs, especially the CVEs. This motivates our RQ-1.2. Besides, from the view of Rust users, they may treat Rust compiler and std-lib as blackbox. Therefore, the robustness of Rust compiler plays an essential role in justifying the effectiveness of Rust and motivates RQ-1.3. 

\uline{\textit{RQ-1.2 How severe are these memory-safety bugs?}}

\begin{figure}[t]
\begin{subfigure}[b]{0.35\textwidth}
\includegraphics[width=0.96\textwidth]{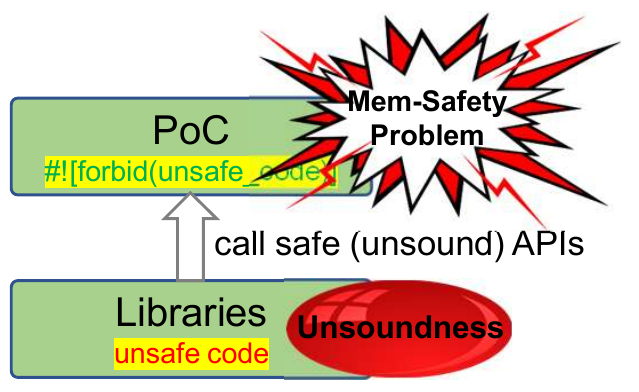}
\caption{Attack model of Rust.}
\label{fig:attackmodel}

\end{subfigure}
\begin{subfigure}[b]{0.56\textwidth}
\includegraphics[width=0.99\textwidth]{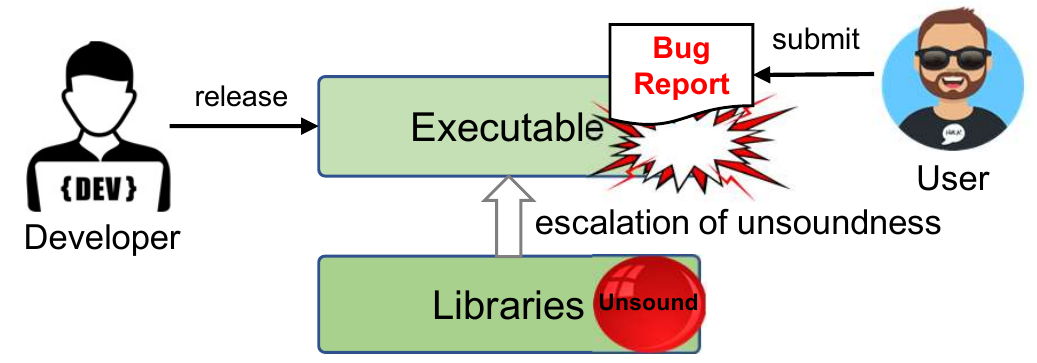}
\caption{Escalation of unsound APIs to real-world bugs.}
\label{fig:escalation}
\end{subfigure}

\caption{The influence of Rust API unsoundness.}
\label{fig:unsound}
\end{figure} 

According to Figure~\ref{fig:dataset}, all existing CVEs are library bugs. This implies such bugs do not cause bad consequences directly unless the buggy APIs are employed by other executables. In comparison, many existing C/C++ CVEs are executable bugs~\cite{fan2020ac}. Furthermore, an important phenomenon of these bugs is that many of them do not contain errors inside but merely introduce unsoundness that violates the memory-safety design of Rust. Consequently, they leave possibilities for their users to write memory-safety bugs without any unsafe marker. For example, all bugs with the culprit of insufficient bound to generic or bad function signature are API declaration issues. When using these APIs carefully, developers will not suffer memory-safety problems. However, since these bugs leave a hole for introducing memory-safety issues without unsafe code, they are also deemed as severe issues and have been assigned CVE-IDs. To demonstrate the existence of such unsoundness issues, bug reporters often need to write complicated proof-of-concept (PoC) samples attacking the vulnerability of unsound APIs. 

Figure~\ref{fig:attackmodel} demonstrates the attack model of Rust. It assumes users should trust all safe APIs provided by libraries that they will not cause memory-safety problems in any circumstances. However, via unsound APIs of libraries, attackers can breach the protection and write memory-safety bugs or PoCs even without unsafe code. Since the consequence and severity of unsound APIs depend on how they are employed, we have further investigated whether these bugs have been propagated to other programs based on each issue raised on GitHub. Figure~\ref{fig:escalation} presents a framework demonstrating how unsound APIs escalate to real-world memory-safety problems. Our investigation result finds no obvious such escalations within the CVEs. The only propagated memory-safety bug is one in our randomly selected program \texttt{curl-rust}\footnote{https://github.com/alexcrichton/curl-rust/issues/333}.

Besides, we also find several other interesting phenomena. Firstly, many bugs are reported by security experts when auditing these Rust projects. For example, \texttt{Qwaz} (GitHub ID) has contributed 29 bugs in our dataset, and \texttt{ammaraskar} has contributed 18 bugs. Secondly, there are several similar bugs due to code clones, such as CVE-2020-36212 copies issue-60977 of std-lib, CVE-2020-36213 copies issue-78498 of std-lib, and issue-21186 of servo copies issue-51780 of std-lib. Thirdly, most project owners are serious about the unsoundness issues within their code while others may not. Taking CVE-2020-35876 as an example, after an unsoundness issue was raised, the project owner closed the issue without a patch but left a few words, ``\textit{Issues like this kill any motivation to improve thing. Not helpful. Open a PR instead}''. 

In short, our analysis reveal that many memory-safety bugs in our dataset are mild, and they only incur possibilities of introducing memory-safety issues without unsafe code. However, since Rust promotes memory-safety, the community is more rigid towards such risks. 

\uline{\textit{RQ-1.3 How robust is Rust compiler?}}

\begin{figure}[t]
\begin{subfigure}[b]{0.58\textwidth}
\includegraphics[width=0.96\textwidth]{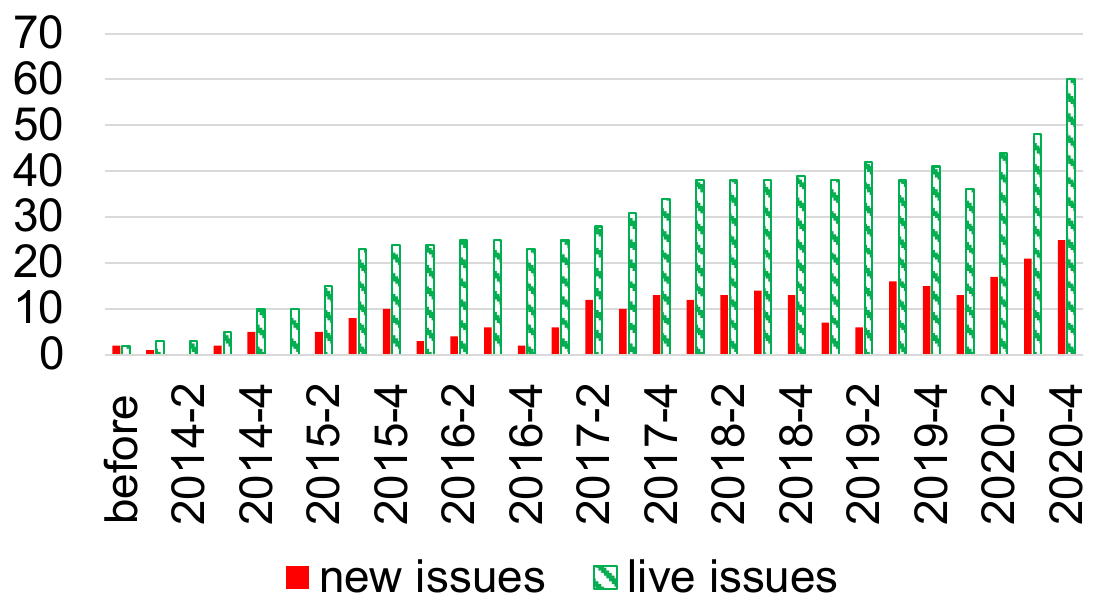}
\caption{Trend of unsoundness issues.}
\label{fig:issue-trend}

\end{subfigure}
\begin{subfigure}[b]{0.4\textwidth}
\includegraphics[width=0.99\textwidth]{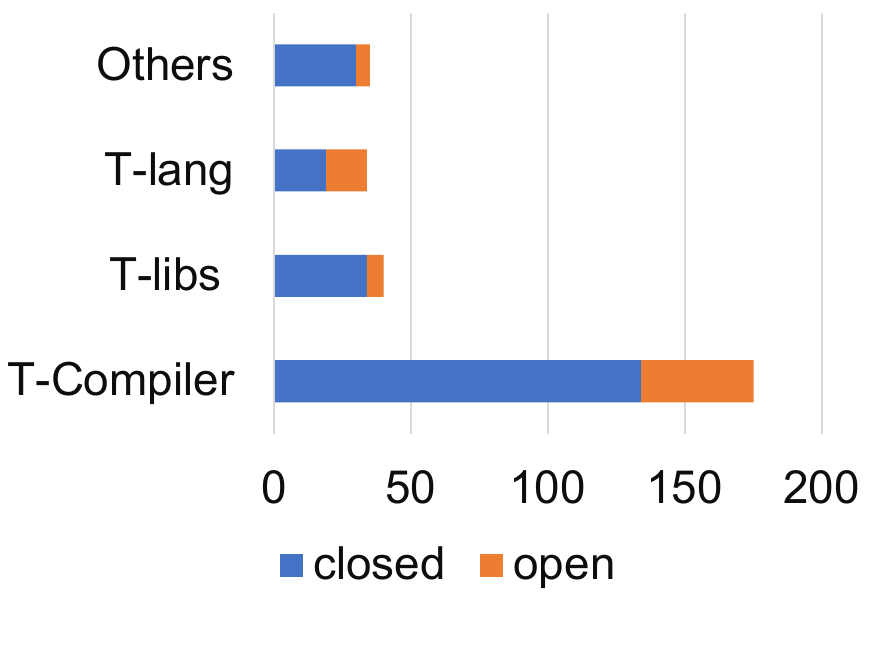}
\caption{Classes of unsoundess issues.}
\label{fig:issue-class}
\end{subfigure}

\caption{Unsoundness issues of the Rust compiler project on GitHub.}
\label{fig:issue-rustc}
\end{figure} 

To answer this question, we have surveyed the unsoundness issues of the Rust compiler project on GitHub. Figure~\ref{fig:issue-trend} presents the number of unsoundness issues raised from 2014 to 2020 in a quarterly manner. Overall, there are over two hundred unsoundness issues (labeled as ``I-unsound'') raised until 2020-12-31, which could violate the memory-safety promise of safe Rust. Most of these issues are also labeled with ``T-Compiler'', which means assigned to the compiler team or compiler bugs. Besides, there are 40 issues labeled with ``T-libs'' (including ``impl-libs'') or ``T-lang'', which means assigned to the library team or language team respectively. We can observe that although the trend of new issues does not decline, it is much stable. Why unsoundness issues can hardly be prevented? As the features of Rust keep getting enriched, introducing new unsoundness issues are unavoidable. Besides, some unsoundness issues related to the design of the language are difficult to fix. Taking CVE-2019-12083 as an example, the temporary fix is to destabilize the function which raises a warning when employing it. However, this fix is imperfect, and the issue still keeps open since raised months ago.

\section{Typical Patterns of Memory-Safety Bugs} \label{sec:patterns}
This section studies the typical patterns of memory-safety bugs and answers RQ-2. Below, we first highlight our main result and then discuss these detailed culprit patterns.

\uline{\textbf{Result Highlight for RQ-2}: We find three major categories of memory-safety bugs typical for Rust: 1) automatic memory reclaim that uncovers the side effect of Rust OBRM which is originally proposed to prevent memory leakage based on static analysis, 2) unsound function that reveals the essential challenge for Rust developers to avoid unsoundness in their code, and 3) unsound generic or trait that unveils another advanced challenge for Rust developers to deliver sound APIs when employing sophisticated language features like polymorphism and inheritance.} 

\subsection{Automatic Memory Reclaim} \label{sec:autoreclaim}

In order to implement OBRM and avoid memory leakage, Rust adopts RAII, \textit{i.e.,} the memory will be automatically acquired while initialization and released if it is no longer used~\cite{ramananandro2012mechanized}. Rust achieves automatic memory reclaim by adding \texttt{drop()} function calls for particular memories in MIR code during compilation. However, since Rust compiler is unsound for unsafe code, such automatic memory reclaim scheme makes Rust programs prone to dangling pointers. In our dataset, 35 bugs are related to such side effect, including 23 bugs occurred during normal program execution and 12 during stack unwinding while an exception is triggered. 

\subsubsection{Bad Drop at Normal Execution Block }
In general, automatic memory reclaim can only trigger memory-safety problems if there is unsafe code, which incurs inconsistencies with Rust OBRM and breaches the soundness of Rust compiler. Below, we discuss several representative scenarios.

\begin{minipage}{\linewidth}
\begin{lstlisting}[language=Rust, caption = PoC of use-after-free and double free bugs due to automatic memory reclaim.\label{list:poc-raii}]
fn genvec() -> Vec<u8> {
   let mut s = String::from("a tmp string");
   /*fix2: let mut s = ManuallyDrop::new(String::from("a tmp string"));*/
   let ptr = s.as_mut_ptr();
   unsafe {
       let v = Vec::from_raw_parts(ptr, s.len(), s.len());
       /*fix1: mem::forget(s);*/
       return v;
       /*s is freed when the function returns*/
   }
}
fn main() {
   let v = genvec();
   assert_eq!('l' as u8,v[0]); /*use-after-free*/
   /*double free: v is released when the function returns*/
}
\end{lstlisting}
\end{minipage}

Typically, automatic memory reclaim would trigger memory-safety problems if the program violates the OBRM model with shared mutable aliases. As a result, dropping one alias would incur dangling pointers with the other one, which may further lead to use-after-free or double free. Code~\ref{list:poc-raii} demonstrates such a self-contained PoC based on CVE-2019-16140. The \texttt{genvec()} function composes a vector using an unsafe API \texttt{from\_raw\_parts()} and returns the vector. The unsafe API simply dereferences the pointer \texttt{ptr} and returns a reference, and it leaves the lifetime checking responsibility to developers. Unfortunately in this PoC, the pointer points to a temporary string \texttt{s} that would be destructed automatically when the function returns. Using the returned vector \texttt{v} in the main function would cause use-after-free issues. Dropping \texttt{v} directly would cause double free. To fix the bug, a common practice is to employ either \texttt{ManuallyDrop<T>} (\textit{e.g.,} by CVE-2018-20996) or \texttt{forget()} ({e.g.,} by CVE-2019-16144) that can prevent calling the destructor of \texttt{T} when it expires. 

In another similar case, shared aliases may not co-exist at the same program point. Instead, the pointer might be used later after the value is dropped, \textit{i.e.,} via usafe APIs or FFIs. CVE-2018-20997 and CVE-2020-35873 are examples of such cases. Note that the lifetime inference scheme associated with OBRM only takes references into consideration but not pointers.

As we have discussed, the problem occurs generally because the constructed variable does not exclusively own its value. Such issues are also popular when designing struct with raw pointers. If the struct does not own the memory pointed by the raw pointer after construction, the memory might be dropped earlier than the struct, \textit{e.g.,} CVE-2020-35711 and CVE-2020-35885. A useful patch to such bug is employing \texttt{PhantomData<T>} which fools the compiler that it owns the data of \texttt{T}. On the contrary, if a struct falsely turns a raw pointer into a reference, it may drop the memory it does not own, \textit{e.g.,} CVE-2020-35885.

The consequences of automatic memory reclaim are generally dangling pointers. However, if the memory has not been fully initialized, it may drop uninitialized memory (\textit{e.g.,} CVE-2020-35888), or incur other undefined behaviors related to the value being dropped (\textit{e.g.,} CVE-2020-35868). 

Note that RAII is not a new concept proposed in Rust. For example, C++ also employs RAII, but its automatic destruction scheme is not as potent as Rust. For comparison, we demonstrate a toy example in Figure~\ref{fig:raii}. Both the two code snippets create an object named \texttt{x} and a pointer that points to another object named \texttt{y}. When the function returns, both \texttt{x} and \texttt{y} are dropped automatically in Rust. However, C++ only drops \texttt{x} automatically, and developers need to delete `y' manually.  

\begin{figure*}[t]
\centering
\begin{subfigure}[b]{0.46\textwidth}
\begin{lstlisting}[language=C++, caption = C++ code: only x is dropped automatically when the function returns.\label{list:raii_c++}]
struct A{
    string name;
    A(string name):name(name){}
    ~A(){cout<<"Drop "<<name<<'\n';}
};
int main()
{
    A x("x");
    A* y = new A("y");
    //delete y;
}

\end{lstlisting}
\end{subfigure}
\hspace{0.5cm}
\begin{subfigure}[b]{0.46\textwidth}
\begin{lstlisting}[language=Rust, caption = Rust code: both x and y are dropped automatically when the function returns.\label{list:raii_rust}]
struct A{name:&'static str}
impl Drop for A {
    fn drop(&mut self) {
        println!("Drop {}", self.name);
    }
}
fn main() {
    let mut x = A{name:"x"};
    let y: *mut A = &mut x;
    unsafe {*y = A{name:"y"}};
}
\end{lstlisting} 
\end{subfigure}
\caption{Comparing the difference of C++ and Rust in automatic drop.}
\label{fig:raii}
\end{figure*}

\subsubsection{Bad Drop at Cleanup Block} \label{sec:dropcleanup}
While the previous type of bugs relates to the side effects of automatic memory reclaim during normal execution, there are also several such bugs only occur during stack unwinding. Rust compiler has two modes of runtime error handling: panic or abort. Panic is the default mode that performs stack unwinding automatically and reclaims the allocated resources; abort simply ends the process directly. This type of bugs only exists in the panic mode.

Memory-safety bugs that occurred during stack unwinding generally imply the existence of program points vulnerable to the cleanup routine, \textit{i.e.,} some memory operations have not been done to achieve a memory-safe state. If the program panics at these points, the cleanup routine may drop dangling or invalid pointers. Code~\ref{list:poc-unwind} present a PoC that would access uninitialized memory if the program panics at line 5 and consequently drops an invalid pointer. Specifically, the function \texttt{read\_from()} creates an object \texttt{foo} with \texttt{uninitialized::<Foo>()} and then initializes it with \texttt{src}. Therefore, the program points after creating \texttt{foo} are vulnerable until \texttt{foo} has been fully initialized. If the program panics at these vulnerable points, the stack unwinding process would call the destructor of \texttt{Foo} which contains an invalid pointer in \texttt{Vec}. Examples of such bugs include CVE-2020-25794 and CVE-2020-36210. The underlying issue is that program points with uninitialized memory are themselves unsound because they violate the principle of RAII. Therefore, Rust introduces a new API \texttt{MaybeUninit} to meet such requirement and deprecated \texttt{uninitialized()} recently. However, there are also other APIs that may also incur similar issues, such as \texttt{alloc()}\footnote{https://gitlab.com/dvshapkin/alg-ds/-/issues/1}. Besides creating new objects, similar problems also exist when expanding a buffer. For example, CVE-2019-16138 expends a vector by firstly increasing its length with \texttt{set\_len()} and then initializing the newly expended elements. In this way, the program points between those two steps are vulnerable to dropping uninitialized memory.

\begin{minipage}{\linewidth}
\begin{lstlisting}[language=Rust, caption = PoC of dropping uninitialized memory during stack unwinding.\label{list:poc-unwind}]
struct Foo{ vec : Vec<i32>, }
impl Foo {
   pub unsafe fn read_from(src: &mut Read) -> Foo {
      let mut foo = mem::uninitialized::<Foo>();
      //panic!(); /*panic here would recalim the uninitialized memory of type <Foo>*/
      let s = slice::from_raw_parts_mut(&mut foo as *mut _ as *mut u8, mem::size_of::<Foo>());
      src.read_exact(s);
      foo
   }
}
fn main(){
   let mut v = vec![0,1,2,3,4,5,6];
   let (p, len, cap) = v.into_raw_parts();
   let mut u = [p as u64, len as _, cap as _];
   let bp: *const u8 = &u[0] as *const u64 as *const _;
   let mut b: &[u8] = unsafe { slice::from_raw_parts(bp, mem::size_of::<u64>()*3) };
   let mut foo = unsafe{Foo::read_from(&mut b as _)};
   println!("foo = {:?}", foo.vec);
}
\end{lstlisting}
\end{minipage}

If the vulnerable program points have shared mutable pointers, panicking at these points would incur double free issues. Note that in Code~\ref{list:poc-raii}, one possible bug fix employs \texttt{forget(s)} to avoid dropping the temporary string \texttt{s}. However, if there are more lines of code that may panic the program after executing \texttt(from\_raw\_parts()) but before reaching \texttt{forget()}, these program points would contain shared mutable pointers to the memory of string \texttt{s}. Panicking at these points would incur double free. Examples of such bugs include CVE-2019-16880 and CVE-2019-16881. A common fix is to replace \texttt{forget()} with \texttt{ManuallyDrop<T>}. Actually, \texttt{forget()} is a wrapper of \texttt{ManuallyDrop<T>} but does not return a \texttt{ManuallyDrop<T>} object. Since \texttt{forget()} borrows the ownership of its parameter value and never returns, it is generally employed as late as possible so that other code can mutate the value. As a result, using \texttt{forget()} is prone to introducing program points vulnerable to stack unwinding, leading to double free or uninitialized memory access (\textit{e.g.,} CVE-2019-15552 and CVE-2019-15553). Besides employing \texttt{forget()} too late, similar problems also arise when shrinking a buffer with pointer elements. For example, if shrinking a vector by firstly copying the elements out and then calling \texttt{set\_len()} to decrease its length, there would be vulnerable program points with duplicated data between the two operations. Such examples include CVE-2018-20991 and CVE-2020-25574.

\subsection{Unsound Function}
An unsound function could be attacked by users and lead to undefined behaviors without unsafe code. This section discusses the issue from two perspectives, bad function signature, and unsoundness introduced by FFIs. Note this category excludes various implementation issues of functions that may also lead to undefined behaviors, as well as wrong API usages that can be fixed easily.

\subsubsection{Bad Function Signature}

According to the composition of a function signature, developers might make mistakes in the following aspects.
\begin{itemize}
\item \textit{Function safety}: If developers falsely declared an unsafe function as safe, users of the function may cause undefined behaviors without using unsafe code themselves. In our dataset, most cases of bad function signature belong to such types, such as CVE-2019-15547, CVE-2019-15548, and CVE-2020-25016.
\item \textit{Parameter mutability}: If developers falsely define an immutable parameter but mutate it internally (\textit{e.g.,} by calling an FFI), users of the function may falsely think the parameter value is not changed and suffer undefined behaviors. Examples of such cases include CVE-2019-16142 and \textit{etc}.
\item \textit{Lifetime bound}: Parameters of functions should be correctly bounded with lifetimes. Missing or insufficient lifetime bound might cause dangling pointers when using these functions, \textit{e.g.,} CVE-2020-35867 and CVE-2020-35879.
\item \textit{Function visibility}: In an extreme case, the whole function is unsound (\textit{e.g.,} deprecated or private) and should not be exposed to users, \textit{e.g.,} CVE-2020-35908.
\end{itemize}

\subsubsection{Unsoundness by FFI}

FFIs are only allowed in unsafe Rust because they could make Rust compiler unsound. To meet the security requirement of Rust, developers should make sure all the behaviors of FFIs should be properly handled or declaring their caller functions as unsafe. Otherwise, the unsoundness of FFIs would propagate to safe Rust code and incur undefined behaviors. 

In our dataset, most cases are due to the undefined behaviors of FFIs with particular applications. For example, issues-17207 of std-lib is unsound because the behavior of \texttt{mallocx()} in jemalloc is undefined for parameter size zero; issue-33770 of std-lib is unsound because the behavior of recursive lock is undefined with glibc. All these undefined behaviors are not properly handled. Moreover, FFIs interacting with the system may suffer race conditions, such as issue-27970 of std-lib and CVE-2020-26235, which are related to \texttt{getenv()} and \texttt{setenv()} of libc. Besides, there could be memory misalignment issues (\textit{e.g.,} CVE-2018-20998 and CVE-2020-35872) or inconsistent data layout (\textit{e.g.,} CVE-2020-35920 and CVE-2020-35921) when interacting with FFIs. Since such bugs are straightforward, we would not demonstrate their trivial details.

\subsection{Unsound Generic and Trait}
Generic and trait are advanced features of Rust type system to support polymorphism and inheritance respectively. While such features can facilitate the development process, it also intensifies the risks of introducing unsoundness. Note that this category only considers the soundness issues of using the generic or trait but excludes various implementation issues of the traits or functions themselves.

\subsubsection{Insufficient Bound of Generic}
In Rust, a function or a struct can take a generic-type parameter as input, \textit{e.g.,} denoted as \texttt{T}, and the specific type of \texttt{T} will be determined during compilation based on how the function is called or how the struct is constructed. However, a practical function may not be able to support all types of variables with any lifetime, and hence bounds are required to make such restrictions. If the specified bounds are insufficient, it may incur memory-safe issues by passing unsupported types.

\begin{minipage}{\linewidth}
\begin{lstlisting}[language=Rust, caption = PoC of lacking \texttt{Send} trait bound to generic.\label{list:poc-tbound}]
struct MyStruct<T> {t:T}
unsafe impl<T> Send for MyStruct<T>{}
//fix: unsafe impl<T:Send> Send for MyStruct<T>{}
fn main(){
    let mut s = MyStruct{t:Rc::new(String::from("untouched data"))};
    for _ in 0..99{
        let mut c = s.clone();
        std::thread::spawn(move ||{
            if !Rc::get_mut(&mut c.t).is_none(){
                (*Rc::get_mut(&mut c.t).unwrap()).clear();
            }
            println!("c.t = {:?}", c.t);
        });
    }
}
\end{lstlisting}

\begin{lstlisting}[language=Rust, caption = PoC of unsound generic that does not respect the memory alignment.\label{list:poc-tgeneric}]
#[repr(align(128))]
struct LargeAlign(u8);
struct MyStruct<T>{ v: Vec<u8>,  _marker: PhantomData<*const T>, }
impl<T:Sized> MyStruct<T>{
    fn from(mut value:T) -> MyStruct<T>{
        let mut v = Vec::with_capacity(size_of::<T>());
        let src:*const T = &value;
        unsafe { 
            ptr::copy(src, v.as_mut_ptr() as _, 1); 
            v.set_len(size)
        }
        MyStruct{ v, _marker:PhantomData }
    }
}
impl<T:Sized> ::std::ops::Deref for MyStruct<T>{
    type Target = T;
    fn deref(&self) -> &T{
        let p = self.v.as_ptr() as *const u8 as *const T;
        unsafe { &*p }
    }
}
fn main(){
    let s = MyStruct::from(LargeAlign(123));
    let v = &*s as *const _ as usize;
    assert!(v % std::mem::align_of::<LargeAlign>() == 0);
}
\end{lstlisting}
\end{minipage}

Code~\ref{list:poc-tbound} presents a PoC that implements the \texttt{Send} trait for \texttt{MyStruct} with a generic type \texttt{T}. However, it misses to bound T with the \texttt{Send} trait. As a result, users are able to construct a \texttt{MyStruct} object with \texttt{RC<T>}. Since \texttt{Rc<T>} is nonatomic, concurrently reading/writing its reference counter would suffer data race. Consequently, the counter value might be incorrect, leading to undefined behaviors when further mutating the wrapped value of \texttt{T}. In this PoC, the spawned thread might wrongly pass the uniqueness check and clear the wrapped string value. If the string buffer is already recycled due to data race, it would cause use-after-free or lead to a malformed stack. Most insufficient bound issues in our dataset are due to lack of \texttt{Send/Sync} bound, such as CVE-2020-35870, CVE-2020-35871, and CVE-2020-35886. The patches of such bugs simply bound \texttt{T} with \texttt{Sync} or \texttt{Send} traits.

Besides \texttt{Sync} or \texttt{Send} traits that are related to concurrent programs, there are also other bounds that may be missed. For example, CVE-2020-35901 and CVE-2020-35902 forget to restrict the generic \texttt{T} to \texttt{Pin<T>}, which means the memory location of \texttt{T} should be pinned; CVE-2020-35906 forgets to bound \texttt{T} with a static lifetime. 

\subsubsection{Generic Vulnerable to Specific Type}
This type of bug also attacks the flexibility enabled by generic, but such bugs cannot be fixed simply by adding bound. Most bugs of this type in our dataset are std-lib bugs when handling special types, such as void (issue-48493) or zero-sized types (issue-42789 and 54857).

Code~\ref{list:poc-tgeneric} presents a PoC based on CVE-2020-25796 and CVE-2020-35903. It defines a struct \texttt{MyStruct} with a generic \texttt{T}. \texttt{MyStruct} contains a constructor \texttt{from()} and an implemented dereference trait. To attack the generic, we define a new struct \texttt{LargeAlign(u8)} with a representation \texttt{\#[repr(align(128))]} that requires 128-byte alignment. Ideally, when dereferencing \texttt{MyStruct} (line 24), the result is a value of type \texttt{LargeAlign} and should be 128-byte aligned. Unfortunately, the current implementation of \texttt{MyStruct} does not respect the alignment and fails the assertion.

\begin{minipage}{\linewidth}
\begin{lstlisting}[language=Rust, caption = PoC of unsound Trait.\label{list:poc-trait}]
trait MyTrait {
    fn type_id(&self) -> TypeId where Self: 'static {
        TypeId::of::<Self>()
    }
}
impl dyn MyTrait {
    pub fn is<T: MyTrait + 'static>(&self) -> bool {
        TypeId::of::<T>() == self.type_id()
    }
    pub fn downcast<T: MyTrait + 'static>(self: Box<Self>) -> Result<Box<T>, Box<dyn MyTrait>>{
        if self.is::<T>(){ unsafe {
           let raw: *mut dyn MyTrait = Box::into_raw(self);
           Ok(Box::from_raw(raw as *mut T))
        }} else { Err(self) }
    }
}
impl<T> MyTrait for Box<T>{}
impl MyTrait for u128{}
impl MyTrait for u8{
    fn type_id(&self) -> TypeId where Self: 'static {
        TypeId::of::<u128>()
    }
}
fn main(){
    let s = Box::new(10u8);
    let r = MyTrait::downcast::<u128>(s);
}
\end{lstlisting}
\end{minipage}

\subsubsection{Unsound Trait}
Traits are abstract classes that can be derived or reimplemented by other types. The feature also leaves substantial room for users to explore attack code, \textit{i.e.,} deriving a safe trait (\textit{e.g.,} ,CVE-2020-25016 and CVE-2020-35866) or reimplementing the trait and causing undefined behaviors (\textit{e.g.,}  CVE-2019-12083 and CVE-2020-25575) without using unsafe code.

Code~\ref{list:poc-trait} demonstrates a PoC based on a std-lib bug CVE-2019-12083. The PoC defines a trait \texttt{MyTrait} with a function \texttt{downcast()} that downcasts the trait to a particular type if \texttt{self.is::<T>()} evaluates to true. \texttt{self.is::<T>()} is based on another function \texttt{type\_id()} which is also implemented by the original trait and always returns a correct type. But \texttt{type\_id()} could be reimplemented by the derived struct and may return a wrong type. As shown in line 21 for example, the reimplemented \texttt{type\_id()} falsely returns a \texttt{u128} type for \texttt{u8}. As a result, it may falsely convert a boxed \texttt{u8} to \texttt{u128} and cause buffer over-read or overflow. Other examples of such type include CVE-2020-25575, CVE-2020-35889, and \textit{etc}.

\section{Lessons Learned} \label{sec:lesson}
Our analysis reveals that a major difference between Rust and other programming languages lies in the soundness promise of APIs. To this end, Rust compiler has already achieved soundness analysis for safe Rust, yet the main challenge lies in unsafe code. This section answers RQ-3 by discussing several suggestions or directions towards mitigating the problem raised by unsafe code.

\uline{\textbf{Result Highlight for RQ-3}: Our lessons learned are two-fold. Firstly, since some bugs demonstrate common patterns, we can summarize several best practices for assisting developers in avoiding unsoundness, such as those for generic bound declaration, avoiding bad drop at cleanup blocks, and function safety declaration. Secondly, since detecting some particular bugs related to unsafe code is not very difficult, we think extending the current static analysis scope of the compiler to support unsafe code should be a promising direction.} 

\subsection{Best Practice for Code Suggestion}
As some bugs in our dataset share common patterns, we can extract these patterns and make useful suggestions to developers for preventing them. 

\subsubsection{Generic Bound Declaration}
The most common bugs in our dataset are those lacking \texttt{Send} or \texttt{Sync} bound to generics when implementing the \texttt{Send} or \texttt{Sync} trait for self-defined structs. 30 bugs in our dataset are of this pattern, such as CVE-2020-35870 and CVE-2020-35886. Fixes of such bugs simply add \texttt{Send} or \texttt{Sync} bound. Therefore, we can recommend developers considering enforcing such bounds when implementing \texttt{Send} or \texttt{Sync} trait. Note that \texttt{Send} or \texttt{Sync} trait are unsafe traits and implementing them are unsafe. If there are other unsafe traits in the future, implementing them may also suffer similar issues.

\subsubsection{Avoiding Bad Drop at Cleanup Block}
As discussed in Section~\ref{sec:dropcleanup}, there are several similar patterns for the bugs related to the bad drop at cleanup blocks, and we can summarize two suggestions based on these patterns.

Firstly, if disabling automatic drop of some variables (\textit{e.g.,} with \texttt{forget()}) is a must, developers should do it as earlier as possible. Otherwise, panicking the program before reaching the disabling statement would be prone to dangling or invalid pointer issues during the stack unwinding process. Examples of such bugs are CVE-2019-15552, CVE-2019-15553, CVE-2019-16880, and CVE-2019-16881. Replacing \texttt{forget()} with \texttt{ManuallyDrop<T>} is sometimes a simple but effective way to fix such bugs, and therefore \texttt{ManuallyDrop<T>} should be recommended to developers as a best practice for such usage. Developers should pay more attention if they insist on using \texttt{forget()}. Because the function borrows the ownership of its parameter value which can prevent further mutating it, it is often impossible to call the function immediately after creating a new alias. This leaves chances to program panic before reaching \texttt{forget()}. Besides, some uninitialized memory issues during stack unwinding can also be prevented by using \texttt{MaybeUninit<T>} instead of the deprecated function \texttt{uninitialized()}. \texttt{MaybeUninit<T>} does not actually allocate the buffer until users have finished the initialization by calling \texttt{assume\_init()}.

Our second suggestion is that when shrinking or expanding a buffer with unsafe code, the length of the resulting buffer should be set at an appropriate program point. In particular, if shrinking a buffer with \texttt{set\_len()}, the \texttt{set\_len()} function should be called as earlier as possible. Calling the function too late (\textit{e.g.,} CVE-2018-20991 and CVE-2020-25574) would lead to shared mutable aliases at some program points. If expending a buffer with \texttt{set\_len()}, the function should be called after the new buffer has been fully initialized, or as late as possible. Otherwise, the new buffer may contain pointers that have not been fully initialized (\textit{e.g.,} CVE-2019-16138), leading to dropping invalid pointers.

\subsubsection{Function Safety Declaration}

\begin{figure}[t!b]
\includegraphics[width=0.96\textwidth]{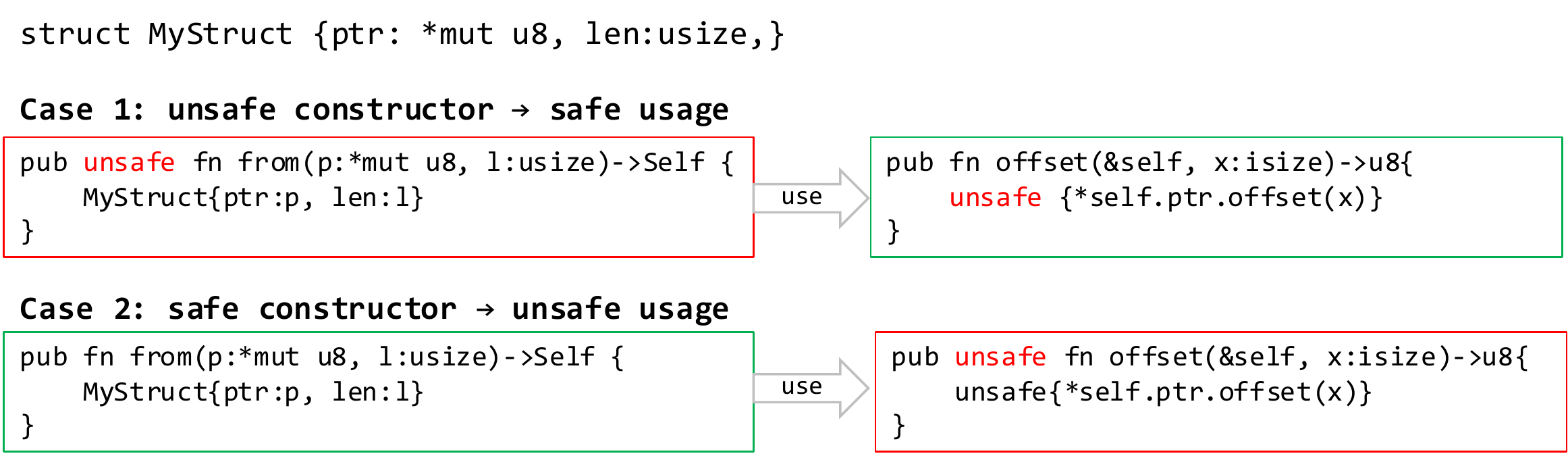}
\caption{Options of function safety declaration for structs with raw pointers.}
\label{fig:mystruct}
\end{figure}

Several unsoundness issues in our dataset falsely declare unsafe functions as safe. As we know, interior unsafe plays an important role in determining the risk of unsoundness~\cite{Evans2020is}. Can we make recommendations to developers based on whether their functions use unsafe code? In general, it is difficult, and our main recommendation is that if a function employs unsafe code unconditionally and it is not a member function, the function is likely to be declared as unsafe. Examples of such bugs that violate the rule include CVE-2019-15547 and CVE-2019-15548. Below we justify the point by discussing the four possible patterns between interior safety and function safety.

\begin{itemize}
\item \textit{unsafe interior $\to$ unsafe function} which means a function that employs unsafe code is declared as unsafe. If the function contains unsafe, the function safety would depend on whether the dangerous inputs have been naturalized before reaching the unsafe code. Without such guarantee, the function should be declared as unsafe by default, such as \texttt{offset()} in Case 2 of Figure~\ref{fig:mystruct}.
\item \textit{unsafe interior $\to$ safe function} which means a function that employs unsafe code is declared as safe. This is the most popular pattern of our function safety declaration bugs. Can we make recommendations based on whether the function employs some restrictions before reaching its unsafe code? \textit{e.g.,} if the first statement is unsafe, does it imply the function should be declared as unsafe? Unfortunately, this is not true, especially for structures. Take \texttt{offset()} in Case 1 of Figure~\ref{fig:mystruct} as an example, it directly dereferences the raw pointer of \texttt{self.ptr}, but it is not a bug because its constructor is declared as unsafe. If the constructor can guarantee that these member values meet safety requirements, then its member functions employ unsafe code directly can also be declared as safe.
\item \textit{safe interior $\to$ safe function} which means a function with pure safe code is defined as safe. If a function does not involve unsafe code internally, it is true that using the function in any way would not cause memory-safety bugs, such as \texttt{from()} in Case 2 of Figure~\ref{fig:mystruct}.
\item \textit{safe interior $\to$ unsafe function} which means a function with pure safe code is defined as unsafe. Although it might be strange, this pattern also exists in practice, such as \texttt{from()} in Case 1 of Figure~\ref{fig:mystruct} or when implementing unsafe traits. 
\end{itemize}

Note that Case 1 and Case 2 of Figure~\ref{fig:mystruct} are two typical patterns for designing structs with raw pointers. Case 1 declares the constructor as unsafe, and therefore all other member functions can be declared safe. On the other hand, case 2 declares the constructor as safe and other member functions as unsafe. Both cases are employed in real-world Rust projects. Based on these cases, we can make another recommendation that if a struct contains members of raw pointers, these members should not be exposed as public to avoid unsound construction by directly assign the raw pointer (\textit{e.g.,} CVE-2020-36205) without using any unsafe marker. Otherwise, all other member functions would be risky to unsoundness.

\subsection{Static Analysis with Unsafe Code}

According to the memory-safety promise of Rust, developers should make sure their provided APIs do not incur undefined behaviors no matter how they are employed. However, preventing general memory-safety bugs via static analysis is hard. According to Rice’s theorem~\cite{rice1953classes}, it is impossible to have a general algorithm for this problem that is sound, complete, and can terminate. Therefore, Rust compiler makes a tradeoff by targeting the soundness analysis of safe code only, leaving the responsibility of employing unsafe code to developers. However, there are also some interesting culprits worth further investigation, especially those related to the side effect of automatic memory reclaim. 

To mitigate the problem of automatic memory reclaim, we may perform alias analysis in the MIR level, and then perform use-after-drop, double drop, or dropping uninitialized memory detection via abstract interpretation~\cite{cousot1996abstract}. Note that although general pointer analysis is hard, this problem can be simplified in two ways. Firstly, considering only the pointers related to unsafe constructors can simplify the analysis problem and should still be effective for most bugs in our dataset. Secondly, there are two approximation directions for solving such problems, the first direction is to perform may-alias-then-check, \textit{i.e.,} to maintain an alias pool recording the may-alias relationship between the source buffer and the newly constructed variable (\textit{e.g.,} via inclusion-based pointer analysis\cite{andersen1994program}) and then check whether a potential bug caused by aliases is spurious. The second direction is to perform precise alias analysis but in a conservative mode, \textit{e.g.,} terminate after some threshold. Taking Code~\ref{list:poc-raii} as an example, the returned value \texttt{v} contains an alias of \texttt{s}, both of which are mutable aliases. If the compiler knows the two pointers could be aligned, it may either report the bug or transfer the ownership from \texttt{s} to \texttt{v} and avoid dropping \texttt{s}. We have done an initial exploration of this problem with a path-sensitive pointer analysis approach and achieved some promising results in both bug recall and overhead. Since its usability would also depend on the precision, a promising direction might be integrating the analysis capability into a specific analysis tool of Rust IDE that is independent to Rust compiler, such as rust-analyzer\footnote{https://rust-analyzer.github.io/} that interacts with the editor via the language server protocol~\cite{meszaros2019delivering}.

Another sophisticated challenge for Rust bug detection is that some APIs are only unsound and they do accomplish bad behaviors by themself. In order to detect these unsoundness issues, one possible direction is to employ automated use case generation based on API dependency graph traversal and then perform analysis based on each use case. Another direction might be leveraging a contract-based approach~\cite{benveniste2007multiple} and detecting whether there are violations of these specified contracts. We leave such investigations as our future work.

\section{Threats to Validity}\label{sec:threat}
Our work performs an in-depth study of memory-safety bugs found in real-world Rust programs. It is mainly based on a dataset of 186 memory-safety bugs. Although the dataset may not be very large, it contains all Rust CVEs with memory-safety issues by 2020-12-31, which are the most severe bugs. Therefore, our analysis result should be representative. Moreover, because our study approach is systematic and in-depth, the dataset is enough for justifying the derived findings and suggestions. As the dataset of bugs may get further enriched in the future, there could be new patterns of culprits found, and the categories employed in Section~\ref{sec:patterns} could be further extended.

\section{Related Work}\label{sec:related}
\subsection{Comparison with Qin's Work} \label{sec:diff}
There are four other independent papers that have similar purposes as we do, including~\cite{yu2019fearless,qin2020understanding,Evans2020is,astrauskas2020programmers}. Both Evans \textit{et al.}~\cite{Evans2020is} and Astrauskas \textit{et al.}~\cite{astrauskas2020programmers} have performed a large-scale study regarding the usage of unsafe code, but without bug analysis. \cite{qin2020understanding} and \cite{yu2019fearless} are produced by the same team, and \cite{yu2019fearless} is a preprint version. Therefore, we mainly compare the difference of our paper with \cite{qin2020understanding}.

The preprint version~\cite{yu2019fearless} by Yu \textit{et al.} studies 18 concurrency bugs from three Rust projects, and \cite{qin2020understanding} by Qin \textit{et al.} extends the work with more bugs and expands the analysis scope from concurrency issues to both memory-safety bugs and concurrency bugs. They categorize 70 memory-safety bugs into four patterns: 1) safe, 2) unsafe, 3) safe->unsafe, and 4) unsafe->safe, and discuss several typical bugs associated with each category. In comparison, our work focuses on memory-safety bugs, and our dataset contains 186 memory-safety bugs which is much larger than \cite{qin2020understanding}. Note that \cite{qin2020understanding} also studies many concurrency bugs, and most of these bugs do not incur memory-safety issues but only deadlock or logical errors that crash the program. Since our work considers more memory-safety bugs, it extracts more bug patterns than \cite{qin2020understanding}, such as the sub-categories of unsound generic or trait. Furthermore, since we focus on memory-safety bugs only, we perform a systematic study regarding these bugs and derive more insights beyond \cite{qin2020understanding}, such as several novel best practices for code suggestion. In short, while \cite{qin2020understanding} provides more understandings of concurrency issues than us, our work provides substantial extensions to the result of \cite{qin2020understanding} in memory-safety bugs.

\subsection{Other Work}
While some work shares experience in using Rust, such as ~\cite{levy2015ownership,anderson2016engineering,levy2017tock,mindermann2018usable}, there are also several papers that focus on the reliability of Rust programs~\cite{toman2015crust,lindner2018no} and the Rust language system~\cite{jung2017rustbelt,baranowski2018verifying,astrauskas2019leveraging}. 

A majority of existing papers on Rust reliability focuses on formal verification~\cite{jung2017rustbelt,dang2019rustbelt,astrauskas2019leveraging} and bug detection~\cite{toman2015crust,dewey2015fuzzing,lindner2018no,baranowski2018verifying}. Formal verification aims at proving the correctness of Rust programs mathematically. RustBelt~\cite{jung2017rustbelt,dang2019rustbelt} is a representative work in this direction. It defines a set of rules to model Rust programs and employs these rules to prove the security of Rust APIs. It has verified that the basic typing rules of Rust are safe, and any closed programs based on well-typed Rust should not exhibit undefined behaviors. Astrauskas \textit{et al.}~\cite{astrauskas2019leveraging} proposed a technique that can assist developers to verify their programs with formal methods. Dewey \textit{et al.}~\cite{dewey2015fuzzing} proposed a fuzzing testing approach to detect the bugs of Rust programs. Lindner \textit{et al.}~\cite{lindner2018no} proposed to detect panic issues of Rust programs via symbolic execution. Besides, some work focuses on the unsafe part of Rust and studies how to isolate unsafe code blocks (\textit{e.g.,}~\cite{lamowski2017sandcrust,almohri2018fidelius,liu2020securing}). We will not go into further details because the purposes of these papers are quite different from our work.

\section{Conclusion}\label{sec:conclusion}
This work studies the effectiveness of Rust in fighting against memory-safety bugs. We have manually analyzed the culprits and consequences of 186 memory-safety bugs covering all Rust CVEs with memory-safety issues. Our study results reveal that Rust successfully limits the risks of memory-safety issues in the realm of unsafe code, and the magic lies in the soundness requirement of safe APIs. As a result, many memory-safety bugs in our dataset are mild soundness issues. Furthermore, we derive three typical categories of memory-safety bugs in Rust, including automatic memory reclaim, unsound function, and unsound generic or trait. In particular, auto memory reclaim uncovers the side effect of Rust OBRM, unsound function unveils the essential challenge of Rust development for avoiding unsoundness, and unsound generic or trait reveals the advanced challenge for developers to ensure soundness when using polymorphism and inheritance. Furthermore, our analysis leads to some suggestions for improving the security of Rust development. For program developers, we provide several recommendations concerning the best practices in using some APIs. For compiler/tool developers or researchers, our suggestion is to consider extending the current program analysis mechanism from safe code to supporting some typical bugs caused by some unsafe code, especially those related to the automatic memory reclaim scheme. We hope our work can raise more discussions regarding the memory-safety effectiveness of Rust and inspire more ideas towards improving the resilience of Rust. 

\section*{Acknowledgement}
We thank all the hunters of bugs employed in our dataset for their contributions in reporting these bugs.

\bibliographystyle{abbrv}
\bibliography{rustbug}

\end{document}